\documentclass[english,floatfix,twocolumn,superscriptaddress,nofootinbib,prd]{revtex4}
\usepackage[utf8]{inputenc}
\usepackage[T1]{fontenc}
\usepackage{lmodern}
\usepackage{amsmath}
\usepackage{amssymb}
\usepackage{graphicx}
\usepackage{esint}
\usepackage{longtable}
\usepackage{dcolumn}
\usepackage{babel} 
\usepackage{csquotes} 
\DeclareMathOperator{\Ima}{Im}

\begin{document}

\title{Thermodynamics of a class of regular black holes with a generalized uncertainty principle}
\author{R. V. Maluf}
\email{r.v.maluf@fisica.ufc.br}
\affiliation{Universidade Federal do Ceará (UFC), Departamento de Física,\\ Campus do Pici, Fortaleza,  CE, C.P. 6030, 60455-760 - Brazil.}

%%%%%%%%%%%%%%%%%%%%%%%%%%%%%%%%%%%%%%%%%%%%%%%%%%%%%%%%%%%%%%%%%%%%%%

\author{Juliano C. S. Neves}
\email{nevesjcs@ime.unicamp.br}
\affiliation{Universidade Estadual de Campinas (UNICAMP), \\ Instituto de Matemática, Estatística e Computação Científica,
 CEP. 13083-859, Campinas, SP, Brazil}

\begin{abstract}
In this article, we present a study on thermodynamics of a class of regular black holes. Such a class includes Bardeen and Hayward regular black holes. We obtained thermodynamic quantities like the Hawking temperature, entropy, and heat capacity for the entire class. As part of an effort to indicate some physical observable to distinguish regular black holes from singular black holes, we suggest that regular black holes are colder than singular black holes. Besides, contrary to the Schwarzschild black hole, that class of regular black holes may be thermodynamically stable. From a generalized uncertainty principle, we also obtained the quantum-corrected thermodynamics for the studied class. Such quantum corrections provide a logarithmic term for the quantum-corrected entropy. 
\end{abstract}

\maketitle

\section{Introduction}

Regular black holes (RBHs) are solutions of gravitational field equations without singularities. It is possible to construct black hole metrics without a physical singularity even in the general relativity (GR) realm. Then, the problem of singularities is solved in a classical theory without a complete quantum theory of gravity. In Bardeen \cite{Bardeen2}, one finds the first RBH defined as a compact object with horizon(s) and without the physical singularity (see \cite{Ansoldi,Lemos_Zanchin} for reviews). Bardeen's RBH is consequence of ideas, of Sakharov and others \cite{Sakharov,Gliner}, that spacetime inside the event horizon for regular solutions is de Sitter-like. The RBH of Bardeen is spherically symmetric and violates the strong energy condition. The violation of an energy condition is the origin for the regularity of Bardeen's black hole. With a de Sitter core, the internal region of regular geometries violates one condition of Hawking-Penrose theorems of singularity \cite{Wald_book}. Then, with such a violation, the existence of a singular point (or a singular ring for axisymmetric spacetimes) is not a necessary consequence from the theorems of singularity. 

Later, other regular solutions with spherical symmetry were proposed by Dymnikova, Bronnikov and Hayward \cite{Dymnikova,Dymnikova2,Dymnikova3,Bronnikov,Hayward}. In particular, the Hayward RBH \cite{Hayward}  was built to describe both the formation and evaporation of black holes. Recently, regular solutions with axial symmetry, or rotation, have been developed \cite{Various_axial,Various_axial2,Various_axial3,Various_axial4,Various_axial5,Neves_Saa} as well. It is shown in \cite{Neves_Saa} that RBHs with rotation always violate the weak energy condition. In \cite{Neves}, in the same direction, we obtained RBHs with spherical symmetry that do not obey the weak energy condition. Regular metrics of black holes were presented in other scenarios, contexts beyond GR like the brane world context \cite{Bronnikov2,Molina_Neves,Neves2}. The class of RBHs investigated in this article was generated by us in an article dedicated to axial symmetry \cite{Neves_Saa}. However, in this paper, we shall focus on RBHs endowed with spherical symmetry. 

Thermodynamics of black holes is one of the most interesting issues in black hole physics today. The notions of temperature and entropy, for example, are associated to black holes as well. Classically, black holes do not emit radiation. However, from a semiclassical point of view, Hawking \cite{Hawking} showed that black holes emit radiation. Today, such a phenomenon is called Hawking radiation, and it is obtained from several approaches. Besides the original Hawking approach, the tunneling method provides a black hole radiation. In such an approach, a particle may cross the event horizon by tunneling. We may indicate two ways to obtain the tunneling result in the literature: the first one is the null-geodesic method developed by \cite{Parikh}; the second one uses the Hamilton-Jacobi ansatz \cite{Agheben} (a brief review is found in \cite{Mann,Mann2}). The latter will be used in this work to evaluate the Hawking temperature of a class of RBHs. Moreover, the black hole temperature is obtained from the first law of thermodynamics as well. According to \cite{Ma}, a corrected first law is necessary  to produce coherent values for the temperature and entropy for RBHs. We adopt the corrected first law proposed by the authors in \cite{Ma}  to evaluate the temperature of a class of RBHs in agreement with a first law that incorporates the so-called Bekenstein-Hawking area law \cite{Bekenstein}, which provides a simple relation between both the entropy $S$ and event horizon area $A$: $S=A/4$.     

Then, with the evaluated temperature, according to thermodynamics, the entropy and heat capacity are obtained. Moreover, it is possible from quantum corrections to calculate the corrected thermodynamic quantities by using a generalized uncertainty principle (GUP).\footnote{See \cite{Tawfik} for a review on GUP and its applications.} Such a generalized principle provides high-energy corrections to black holes thermodynamics, which come from a supposed quantum theory of gravity and the idea of a minimum length. For the class of RBHs studied here, the quantum correction for entropy is also logarithmic---in agreement with Kaul and Majumdar in the pioneering work \cite{Kaul}. The GUP has been applied in several black holes metrics. Besides the standard black holes in GR \cite{Various_GUP,Various_GUP2,Various_GUP3}, the GUP was applied in self-dual black holes \cite{Anacleto}, brane world black-holes \cite{Casadio}, Horava-Lifshitz black holes \cite{Anacleto2}, and others. 

Thermodynamics of RBHs has been studied in several articles \cite{Thermodynamics,Thermodynamics2,Thermodynamics3,Thermodynamics4,Thermodynamics5,Thermodynamics6,Thermodynamics7,Tharanath,Sharif}. A thermodynamic observable may provide a form to distinguish RBHs from singular black holes. From three different ways---surface gravity, the tunneling method and the first law of thermodynamics---we obtained Hawking temperature of a class of RBHs. The corresponding temperature of the class indicates that \textit{regular black holes are colder than the Schwarzschild black hole}. Moreover, contrary to the Schwarzschild black hole, it is shown that the class of RBHs may be thermodynamically stable, i.e., its corresponding heat capacity may be positive. 

The structure of this paper is as follows: in Sec. II we present the class of black holes developed in \cite{Neves_Saa} focusing on spherical solutions. In Sec. III, we show the thermodynamics of such a class. The quantum-corrected thermodynamics of RBHs within that class is indicated using the GUP in Sec. IV. The final remarks are presented in Sec V.  We adopt Planck units, in which the speed of light in vacuum $c$, the gravitational constant $G$, Boltzmann constant $k_B$, and the reduced Planck constant $\hslash$ are set equal to 1.

\section{A class of regular black holes}

Considering metrics with spherical symmetry, it is possible to build, from the general metric in the ($t,r,\theta,\phi$) coordinates
\begin{equation}
ds^2=-f(r)dt^2+\frac{dr^2}{f(r)}+r^2 \left( d\theta^2+\sin^2\theta d\phi^2 \right),
\label{Metric}
\end{equation}
regular solutions of black holes, or RBHs, by assuming the following form for the metric term $f(r)$: 
\begin{equation}
f(r)=1-\frac{2m(r)}{r}.
\end{equation}
The general mass function presented in \cite{Neves_Saa} is given by
\begin{equation}
m(r)=\frac{M_0}{\left[1+\left(\frac{r_0}{r}\right)^{q} \right]^{\frac{p}{q}}},
\label{Mass}
\end{equation}
where $M_{0}$ and $r_{0}$ may be interpreted as mass and length parameters, respectively. The parameter $M_0$ stands for the Arnowitt-Deser-Misner (ADM) mass of the Schwarzschild black hole. The limit of $m(r)$ confirms such an interpretation, i.e., $\lim_{r\rightarrow\infty}m(r)=M_0 $. Indeed, the metric (\ref{Metric}) with the mass function $m(r)$ is approximately the Schwarzschild metric for large values of the radial coordinate. On the other hand, for small scales, $r_0$ may be seen as a microscopical length. Thus, without such a microscopical length ($r_0=0$), the metric (\ref{Metric}) with the mass function given by Eq. (\ref{Mass}) is the Schwarzschild metric. The values of $p$ and $q$ are integer and positive. To obtain the so-called Bardeen or Hayward RBHs, one specifies $p=3$ and $q=2$ or $p=q=3$ in the general mass function, respectively. The existence of horizon(s) (at most two in Bardeen's solution, the internal $r_-$ and the external $r_+$ horizons) is provided by $r_0 < M_0$.  As we can see, the mass function (\ref{Mass}) provides a class of RBHs. 

Then, the basic properties indicated by the mass function given by Eq. (\ref{Mass}) are: (i) a de Sitter core is generated for small values of $r$ since $p=3$, as indicated in \cite{Neves_Saa}. Therefore, the appropriate value of $p$ for generating regular solutions is 3. For $p=3$, the metric term at small scales is $f(r)\sim 1-Cr^2$, with $C$ playing the role of a positive cosmological constant, because in these scales $m(r) \sim r^3$. (ii) for large values of $r$ in Eq. (\ref{Mass}), typically $r \gtrsim r_+$, the mass function is almost constant ($m(r)\approx M_0$), then we have approximately Schwarzschild metric. In particular, independently of $r$, the Schwarzschild metric is restored  for $r_0=0$. Thus, the regular class with spherical symmetry generated by (\ref{Mass}) has its singular counterpart given by the Schwarzschild metric. (iii) metrics generated by $m(r)$ are not vacuum solutions of Einstein's field equations ($T_{\mu\nu} \neq 0$). But according to the shape of $m(r)$, for $r>r_+$ the class is described almost by vacuum solutions. 

As a \textit{good} RBH solution, the regular spacetimes generated by (\ref{Mass}) provide bounded scalars. All RBHs of the studied class have nonsingular scalars. As we said, the existence of a de Sitter core avoids the appearance of a singular point, which is translated by the divergence of scalars in that point. For example, for the studied class, the Kretschmann scalar $K(r)$, given by Riemann tensor $R^{\mu}_{\ \nu \alpha \beta}$, has its limit written as
\begin{equation}
\lim_{r\rightarrow 0} K(r) = \lim_{r\rightarrow 0} R_{\mu\nu\alpha\beta}R^{\mu\nu\alpha\beta}=96\left(\frac{M_{0}}{r_{0}^3}\right)^2
\label{K_Bardeen}
\end{equation}
at origin. As we can see, independently of $q$, the entire class of black holes given by (\ref{Mass}) is regular.      

The singularity theorems state the existence of singularities in spacetimes using the concept of geodesics (see Wald's book \cite{Wald_book}, chapter 9). A solution of Einstein's field equations, like Schwarzschild or Friedmann-Lemaître-Robertson-Walker, presents a singularity if it is geodesically incomplete. But the standard solutions in GR are geodesically incomplete and have an unbounded scalar at same time. For such solutions, unbounded scalars are primary indications of the existence of singularities. On the other hand, there are exotic geometries that are not geodesically incomplete, but such spacetimes have problems with scalars \cite{Olmo}. 

\section{Thermodynamics}
In this section, we study thermodynamic quantities without quantum corrections. For the class of regular metrics generated by the mass function (\ref{Mass}), we present its corresponding temperature, entropy, and heat capacity. For temperature, it is shown three forms to evaluate it: by means of surface gravity, $\kappa$, the tunneling method and by the first law of thermodynamics. As we shall see, the first law of black hole thermodynamics needs a correction for RBHs. In \cite{Ma} such a corrected term is shown, and we apply it to the class analyzed here. Regarding the entropy, the corrected first law provides an appropriate value for the RBHs entropy: $A/4$, where $A$ is the event horizon area. Contrary to the Schwarzschild black hole, it is possible to build thermodynamically stable RBHs, i.e., as we shall see, it is possible to obtain positive values for the heat capacity for the class of RBHs considered in this article.

\subsection{Temperature}
The metric given by Eq. (\ref{Metric}) is endowed with a timelike Killing vector $\xi=\partial/\partial t$. Thus, it has a conserved quantity associated to $\xi$. It is possible to construct a conserved quantity by using that Killing vector such that
\begin{equation}
\nabla^{\nu} \left(\xi^\mu \xi_\mu \right) = -2 \kappa \xi^{\nu}, 
\end{equation} 
where $\nabla^{\nu}$ is the covariant derivative and $\kappa$ is constant along $\xi$ orbits, i.e., the Lie derivative of $\kappa$ along $\xi$ vanishes:
\begin{equation}
\mathcal{L}_\xi \kappa =0.
\end{equation} 
In particular, $\kappa$ is constant over the horizon, and it is called surface gravity. In the coordinate basis, one reads $\xi ^\mu = (1,0,0,0)$ for the timelike Killing vector components, and the surface gravity for the metric ansatz (\ref{Metric}) is written as
\begin{equation}
\kappa=\frac{f'(r)}{2} \Big\lvert_{r_+},
\end{equation}
with  $'$ denoting a derivative with respect to the radial coordinate. Hawking showed \cite{Hawking} that black holes emit radiation, and its corresponding temperature---Hawking temperature---is given by
\begin{equation}
T_\kappa =\frac{\kappa}{2\pi},
\label{T_kappa}
\end{equation}
for stationary spacetimes. Regarding the metric (\ref{Metric}) and the mass function (\ref{Mass}), it is straightforward to show that the above definition provides
\begin{equation}
T_\kappa = \frac{1}{4\pi r_{+}} \left[1-2\left(\frac{r_0}{r_+} \right)^q \right] \left[ 1+\left(\frac{r_0}{r_+}\right)^q\right]^{-1}
\label{T_kappa2}
\end{equation}
for the entire class of RBHs. To obtain the temperature (\ref{T_kappa2}), we used the relation
\begin{equation}
M_0=\frac{r_+}{2}\left[1+\left(\frac{r_0}{r_+}\right)^{q} \right]^{\frac{3}{q}},
\label{M0}
\end{equation}
which comes from the metric term equation $f(r_+)=0$. As we can see, the Schwarzschild temperature is obtained by making $r_0=0$ ($T_{Sch}=1/4\pi r_+$, assuming that  $M_0$ is the Schwarzschild mass, i.e., $r_+ = 2M_0$).  

Another form to calculate the black hole temperature is given by the tunneling effect. The quantum tunneling effect allows that particles inside the black hole cross the event horizon. It is possible to calculate the tunneling probability of this process, as indicated in \cite{Agheben,Mann,Mann2}. In such an approach, we are interested in radial trajectories, then the metric studied may be considered two dimensional near the horizon:
\begin{equation}
ds^2=-f(r)dt^2+\frac{dr^2}{f(r)}.
\label{metric-2d}
\end{equation}
Thus, the problem is entirely solved in the $t-r$ plane. The Klein-Gordon equation for a scalar field $\phi$ with mass $m_{\phi}$ is
\begin{equation}
\hslash^2 g^{\mu\nu}\nabla_{\mu}\nabla_{\nu}\phi-m_{\phi}^2\phi=0,
\end{equation}
and, with the aid of Eq. (\ref{metric-2d}), is written as
\begin{equation}
-\partial_{t}^{2}\phi +f(r)^2 \partial_{r}^{2}\phi+\frac{1}{2}\partial_{r}f(r)^2\partial_{r}\phi-\frac{m_{\phi}^2}{\hslash}f(r)\phi=0.
\label{equation}
\end{equation}
Using the Wentzel-Kramers-Brillouin (WKB) method, one has the following solution for Eq. (\ref{equation}):
\begin{equation}
\phi (t,r)= \exp \left[-\frac{i}{\hslash}I(t,r) \right].
\end{equation}
For the lowest order in $\hbar$, one has the Hamilton-Jacobi equation
\begin{equation}
\left(\partial_{t}I \right)^2 -f(r)^2 \left(\partial_{r}I \right)^2-m^2f(r)=0,
\label{HJ}
\end{equation}
with 
\begin{equation}
I(t,r)=-Et+W(r)
\label{Action}
\end{equation}
playing the role of the action that generates (\ref{HJ}), and $E$ is the radiation energy measured, for example, outside the event horizon. The explicit form for $W(r)$, the spatial part of the action, reads 
\begin{equation}
W_{\pm}(r)=\pm \int\frac{dr}{f(r)}\sqrt{E^2-m_{\phi}^2f(r)}.
\label{W}
\end{equation}
The functions $W_{\pm}(r)$ represent outgoing and ingoing solutions, respectively. Classically, $W_{+}(r)$ is forbidden because it describes solutions that cross the event horizon, moving away from $r_+$. Then, to obtain Hawking radiation outside the event horizon, we shall focus on $W_{+}(r)$. 

With the approximation for the function $f(r)$ near the event horizon $r_+$, i.e.,
\begin{equation}
f(r)= f(r_+)+f'(r_+)(r-r_+)+...,
\end{equation}
 Eq. (\ref{W}) assumes the simple form
\begin{equation}
W_{+}(r)=\frac{2\pi i E}{f'(r_+)}.
\end{equation}

Therefore, for a particle, the tunneling probability of crossing the event horizon is given by the imaginary part of (\ref{Action}):
\begin{equation}
\Gamma \simeq \exp \left[-2 \Ima I \right] = \exp\left[-\frac{4 \pi E}{f'(r_+)} \right].
\label{Gamma}
\end{equation}
Comparing Eq. (\ref{Gamma}) with the Boltzmann factor $e^{-E/T}$, Hawking temperature derived by the tunneling method is written as
\begin{equation}
T_t=\frac{E}{2\Ima I}=\frac{f'(r_+)}{4\pi}=T_\kappa.
\end{equation} 
As we can see, both the temperature from Hawking's initial idea, $T_\kappa$, and the temperature by tunneling method, $T_t$, are in agreement. 

The third form to calculate black hole temperature is provided by the first law of thermodynamics. For vacuum spacetimes with spherical symmetry, the first law of black holes thermodynamics says that
\begin{equation}
dM=T_f dS,
\label{First_law}
\end{equation}
where $M$ and  $S$ are the total energy and entropy of the system, respectively, and $T_f$ is the black hole temperature provided by the first law. According to \cite{Ma}, assuming the area law ($S=A/4=\pi r_{+}^2$) and $M=M_0$, the first law does not provide a correct form for RBHs temperature. Explicitly, Eq. (\ref{First_law}) leads to
\begin{align}
T_f = \frac{dM}{dS}&= \frac{1}{2\pi r_+}\frac{dM_0}{dr_+} \nonumber \\ 
&=\frac{1}{4\pi r_+} \left[1-2\left(\frac{r_0}{r_+} \right)^q \right]  \left[ 1+\left(\frac{r_0}{r_+}\right)^q\right]^{\frac{3-q}{q}},
\label{T_not}
\end{align}
using the mass function (\ref{Mass}). Then, as we can see, $T_\kappa$ and $T_f$, given by Eqs. (\ref{T_kappa2}) and (\ref{T_not}), respectively, are not in agreement. That is, the first law given by (\ref{First_law}) is not appropriate for calculating the correct black hole temperature for RBHs. But, as we said, solutions generated by $m(r)$ are not vacuum solutions. In \cite{Sharif}, where thermodynamics of Bardeen RBH was studied, the indicated alternative was to add an extra term in Eq. (\ref{First_law}), given by the supposed electromagnetic origin of Bardeen's solution (see in \cite{Beato} Ayon-Beato and Garcia's interpretation of a Bardeen RBH as a solution of GR coupled to a nonlinear electrodynamics). Then, the first law is written as 
\begin{equation}
dM=TdS+\Phi de,
\label{First_law2}
\end{equation}
with $e=r_0$ playing the role of a magnetic monopole in the interpretation of the Bardeen RBH suggested in \cite{Beato}, $M=M_0$,  and the potential $\Phi$ is given by
\begin{equation}
 \Phi =\frac{ \partial M}{\partial e} \Big\vert _{r=r_+}.
\end{equation}
However, from $T=T_\kappa$, or Hawking temperature calculated by surface gravity, the authors do not obtain the correct form for the entropy, or area law, using the first law (\ref{First_law2}).

We saw two different methods to calculate the black hole temperature that provide the same result: by means of the surface gravity and tunneling. However, using the first law (\ref{First_law}), the same result is not obtained. On the other hand, the first law (\ref{First_law2}) assumes the correct result for temperature but does not provide the area law for entropy. Thus, in our perspective, the third way is not correct for RBHs, and an adjustment is necessary. Then, following \cite{Ma} and assuming the validity of the area law, we find a corrected first law of thermodynamics that fixes the temperature for the entire class of RBHs studied in this article. The corrected temperature is obtained from the following first law:
\begin{equation}
F(r_+,r_0)dM_0=T_{f}' dS.
\label{First_corr}
\end{equation}
The factor $F(r_+,r_0)$, which depends on the terms of mass function (\ref{Mass}), fixes the first law for RBHs for several classes (not only for the class presented here). Among the RBHs included by the corrected first law, our class and the noncommutative RBHs \cite{Nicolini} are present. According to \cite{Ma}, the correction is defined as
\begin{equation}
F(r_+,r_0) = 1+4\pi \int_{r_+}^{\infty}r^2\frac{\partial T_{0}^{0}}{\partial M_0}dr.
\end{equation}
$T_{0}^{0}$ is the energy-momentum component that corresponds to the energy density. For metrics where the energy density does not depend on the mass, the correction vanishes. Using our class of RBHs, one has its explicit form
\begin{equation}
F(r_+,r_0)=\left[ 1+\left(\frac{r_0}{r_+} \right)^q \right]^{-\frac{3}{q}},
\label{Correction}
\end{equation}   
which is just the factor that defines the mass function (\ref{Mass}). From the correction $F(r_+,r_0)$, it is straightforward to obtain a corrected form for temperature ($T_{f}'$) using (\ref{First_corr}), which is in agreement with the other two different approaches:
\begin{equation}
T_{f}' = T_{\kappa} = T_{t} = F(r_+,r_0)T_{f}.
\end{equation} 
Thus, with a correct first law, we have both the correct temperature and entropy. 

With a reliable form for the temperature of the RBHs class, it is interesting to compare the Schwarzschild temperature with the corresponding thermodynamic quantity for RBHs. It is more convenient to write the mass function for large values of $r$ ($r_0/r \ll 1$, i.e., for $r \gtrsim r_+$) as
\begin{equation}
m(r) \approx M_0 \left( 1-\frac{3}{q}\left(\frac{r_0}{r}\right)^q  \right).
\label{Approx}
\end{equation}
This approximation leads to a simplified result for Hawking temperature and, consequently, a clear comparison to Schwarzschild black hole. Using any method illustrated above to evaluate the temperature, one has an approximate temperature using the mass function (\ref{Approx}) 
\begin{equation}
 T_{app} \approx T_{Sch} \left[1-\frac{3 \left(1+q \right)}{q} \left(\frac{r_0}{r_{+}}\right)^q \right],
\label{Approximate}
\end{equation}
assuming that $r_+$ is approximately the Schwarzschild radius. As we can see, the approximate temperature is the temperature of Schwarzschild black hole minus a positive term from the regular metric. That is, \textit{for the class studied in this work, RBHs are colder than the Schwarzschild black hole}. This interesting feature of RBHs may be considered an observational discrepancy between regular and singular black holes. Recently, several authors have proposed ways to distinguish RBHs from observations \cite{applications,applications2,applications3,applications4,applications5,applications6,applications7,applications8,applications9,applications10}. For us, Hawking temperature may be a form to achieve such an observational test (even knowing that Hawking temperature is tiny for large black holes). But it is interesting to consider that a thermodynamic quantity may be able to provide some information on the interior region of black holes. In this case, the absence or presence of a physical singularity.    

\subsection{Entropy}
As we said, the area law is assumed and satisfied in this article, i.e., for the entire class of RBHs, entropy is $S=A/4$. With the corrected first law given by Eq. (\ref{First_corr}), we have the verification of the area law: 
\begin{equation}
S = \int \frac{F(r_+,r_0)}{T_{f}'}dM_0=\pi r_{+}^2=\frac{A}{4},
\label{Entropy}
\end{equation}
which is straightforward using Eqs. (\ref{T_kappa2}), (\ref{M0}), and (\ref{Correction}). 

\begin{center}
\begin{figure}
\includegraphics[scale=0.487]{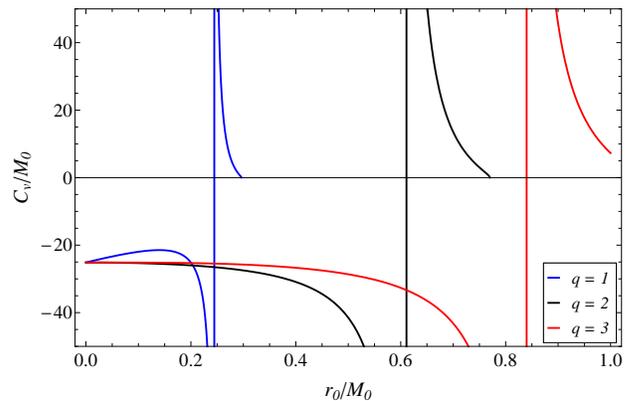}
\caption{Heat capacity at constant volume, $C_v$. The figure illustrates possible positive values for $C_v$ for members of the class of RBHs. Positive values of the heat capacity indicate thermodynamic stability. $q=2$ and $q=3$ describe the so-called Bardeen and Hayward black holes, respectively. The discontinuity indicates phase transition. In this graphic $M_0=1$, and we assume $r_0<M_0$ to obtain a structure of horizons for the regular metrics. For $r_0=0$, the heat capacity assumes the corresponding value for Schwarzschild's black hole: $C_{v(Sch)}=-2\pi r_{+}^{2}=-8\pi M_0^2$, which in this graphic is approximately $-25.1$.}
\end{figure}
\end{center}

\subsection{Heat capacity}
The heat capacity gives the information on the thermodynamic stability of a system. Negative heat capacity indicates an unstable system. It  is known that the Schwarzschild black hole is unstable from thermodynamic considerations. In Schwarzschild, the heat capacity at constant volume is $C_{v(Sch)}=-2\pi r_{+}^2$. According to Wald \cite{Wald}, the reason for the instability is in the infinite thermal bath of Schwarzschild's system. But for the class of RBHs given by the mass function (\ref{Mass}), it is possible to obtain thermodynamically stable black holes from a suitable choice of parameters. The heat capacity at constant volume is defined as
\begin{equation}
C_v = T \frac{\partial S}{\partial T} = T \frac{\partial S}{\partial r_+} \left(\frac{\partial T}{\partial r_+} \right)^{-1}, 
\label{HC_def}
\end{equation}    
where $T$ is the temperature given by any method illustrated above. For the class studied here, it is written as
\begin{equation}
C_v = -2\pi r_{+}^2 \left(1+ \frac{3q \left(\frac{r_0}{r_{+}}\right)^q}{1- \left(1+3q\right) \left(\frac{r_0}{r_{+}}\right)^q-2\left(\frac{r_0}{r_{+}}\right)^{2q}} \right).
\label{Heat}
\end{equation}
As we can see, when  $r_0=0$, Eq. (\ref{Heat}) is just the heat capacity of the Schwarzschild black hole $C_{v(Sch)}=-2\pi r_{+}^2$. However, positive values for $C_v$ are possible when
\begin{equation}
\left( \frac{-1-3q+ 3 \sqrt{q^2+\frac{6q}{9}+1 }}{4}  \right)^{\frac{1}{q}}r_+ < r_0 < \frac{r_+}{2^{\frac{1}{q}}}. 
\label{Relation}
\end{equation}
For this range of values for the length parameter $r_0$, the heat capacity is positive (see Fig. 1), and the entire class of RBHs is thermodynamically stable. Besides stability, the horizon structure is possible within this range of values, i.e., as we said, the existence of horizons ($r_-$ and $r_+$) is assured by $r_0<M_0$. According to \cite{Tharanath}, where Bardeen and Hayward geometries were studied, the discontinuity in the heat capacity for RBHs is interpreted as a phase transition.  

\section{Quantum corrections to RBHs thermodynamics}

To obtain the quantum corrections for the thermodynamic quantities shown above, we shall consider a generalized uncertainty principle (GUP) applied to tunneling formalism. This procedure has been used to calculate the quantum corrections for the temperature, entropy, and heat capacity for different types of black holes \cite{Various_GUP,Various_GUP2,Various_GUP3,Anacleto,Casadio,Anacleto2}. In this approach, the quadratic GUP is written as
\begin{equation}
\triangle x\triangle p\geq\hslash\left(1+\frac{\lambda^{2}l_{p}^{2}}{\hslash^{2}}\triangle p^{2}\right),
\label{GUP}
\end{equation}
where $\lambda$ is the so-called dimensionless quantum gravity parameter, and $l_{p}=\sqrt{\frac{\hslash G}{c^{3}}}\approx10^{-35}m$ is the Planck length. We can rewrite the GUP (\ref{GUP}) as
\begin{equation}
\triangle p\geq\frac{\hslash \triangle x}{2\lambda^{2}l_{p}^{2}}\left(1-\sqrt{1-\frac{4\lambda^{2}l_{p}^{2}}{\triangle x^2}}\right),
\end{equation}
and considering $l_p/\triangle x \ll 1$, we apply Taylor expansion such that
\begin{equation}
\triangle p\geq\frac{1}{\triangle x}\left(1+\frac{2\lambda^{2}l_{p}^{2}}{\triangle x^{2}}+\cdots\right),
\end{equation}
where $\hslash$ was set equal to 1, according to Planck units.

Using the saturated form of the uncertainty principle, namely $E\triangle x\geq1$, the energy correction reads 
\begin{equation}
E_{GUP}\geq E\left(1+\frac{2\lambda^{2}l_{p}^{2}}{\triangle x^{2}}\right),
\end{equation}
up to second order in $l_{p}$. Therefore, according to Eq. (\ref{Gamma}), for particles with corrected energy $E_{GUP}$, the tunneling probability of crossing the event horizon is
\begin{equation}
\Gamma \simeq \exp \left[-\frac{4\pi E_{GUP}}{f'(r_{+})}\right]=\exp \left[-\frac{E_{GUP}}{T_{\kappa}}\right].
\label{Gamma_GUP}
\end{equation}
Thus, comparing  Eq. (\ref{Gamma_GUP}) with the Boltzmann distribution $e^{-E/T}$, we found the quantum-corrected Hawking temperature:
\begin{equation}
T_{GUP}=T_{\kappa}\left(1+\frac{\lambda^{2}l_{p}^{2}}{2r_{+}^{2}}\right)^{-1},
\label{T_corrected}
\end{equation}
where $T_{\kappa}$ is given by Eq. (\ref{T_kappa2}) and, according to \cite{Vagenas}, the uncertainty in $x$ for events near the event horizon is $\triangle x\simeq 2r_{+}$. The quantum correction for the black hole entropy is straightforwardly obtained by the relation
\begin{equation}
S_{GUP}=\int\frac{F(r_{+},r_{0})}{T_{GUP}}dM_{0},
\end{equation}
where the integral leads to
\begin{equation}
S_{GUP}=\pi\left(r_{+}^{2}+\lambda^{2}l_{p}^{2}\ln r_{+}\right).
\end{equation}
As we can see, the quantum correction for entropy presents the so-called logarithmic term, as indicated by Kaul and Majumdar \cite{Kaul}. But such a term is very tiny using the GUP (\ref{GUP}). According to recent studies \cite{Das,Vagenas2}, the upper bound for the quantum gravity parameter is $\lambda \sim 10^{10}$. Then, adopting that upper bound, we have a factor $10^{-50}$ multiplying the logarithmic term.     

Using the quantum-corrected temperature and entropy, the correction to the heat capacity, which is defined in (\ref{HC_def}), is  given by
\begin{equation}
(C_{v})_{GUP} = T_{GUP}\frac{\partial S_{GUP}}{\partial T_{GUP}}.
\end{equation}
Then, for the entire class of RBHs, one has
\begin{widetext}
\begin{equation}
(C_{v})_{GUP} =\frac{\pi\left(2r_{+}^{2}+\lambda^{2}l_{p}^{2}\right){}^{2}\left[1+\left(\frac{r_{0}}{r_{+}}\right)^{q}\right]\left[-1+2\left(\frac{r_{0}}{r_{+}}\right)^{q}\right]}{2r_{+}^{2}-\lambda^{2}l_{p}^{2}-\left[2(1+3q)r_{+}^{2}+\lambda^{2}l_{p}^{2}(-1+3q)\right]\left(\frac{r_{0}}{r_{+}}\right)^{q}-2\left(2r_{+}^{2}-\lambda^{2}l_{p}^{2}\right)\left(\frac{r_{0}}{r_{+}}\right)^{2q}}.
\end{equation}
\end{widetext}
Note that when $\lambda\rightarrow 0$, we recover the previous result:
\begin{equation}
\lim_{\lambda\rightarrow 0}(C_{v})_{GUP}=C_{v},
\end{equation}
 where $C_{v}$ is given by Eq. \eqref{Heat}. And then assuming $r_0 \rightarrow 0$, we have the heat capacity of the Schwarzschild black hole.

\section{Final remarks}
In this article, we presented a study on thermodynamics of a class of regular black holes (RBHs). The class contains, for example, the so-called Bardeen and Hayward RBHs. Hawking temperature was obtained from three forms. Following \cite{Ma}, from the first law of thermodynamics, we needed to assume  a term that corrects such a law and provides a correct result for temperature and, consequently, assures a first law in agreement with the entropy, or Bekenstein-Hawking area law. Then, from the correct results for the temperature and entropy, the heat capacity at a constant volume was obtained, indicating the thermodynamic phase transition and stability for the class of RBHs. That is, regarding stability, contrary to the Schwarzschild black hole, it is possible to build stable thermodynamic solutions within that class of RBHs.

It is worth to emphasize another difference between the class of RBHs and the Schwarzschild black hole. The temperature of RBHs indicates that the members of the studied class are colder than the Schwarzschild black hole. This may be argued as an observable difference between both regular and singular black holes.

Lastly, quantum corrections to thermodynamic quantities were analyzed. In particular, quantum corrections from a generalized uncertainty principle (GUP) provided a tiny logarithmic correction to the entropy of the class of RBHs in agreement with several works. 

\section*{Acknowledgments}
RVM would like to thank Conselho Nacional de Desenvolvimento Científico e Tecnológico (CNPq) for financial support (Grant No. 305678/2015-9). JCSN would like to thank Department of Physics, at Universidade Federal do Ceará, for the kind hospitality.


\begin{thebibliography}{99}

\bibitem{Bardeen2}J. M. Bardeen, in\textit{Conference Proceedings of GR5} (Tbilisi, URSS, 1968), p. 174.

\bibitem{Ansoldi}S. Ansoldi, \textit{Spherical black holes with regular center: a review of existing models including a recent realization with Gaussian sources} [arXiv:0802.0330].

\bibitem{Lemos_Zanchin}J. P. S. Lemos and V. T. Zanchin, Phys. Rev. D \textbf{83}, 124005 (2011).

\bibitem{Sakharov}A. D. Sakharov, \textit{Sov. Phys. JETP} \textbf{22}, 241 (1966).

\bibitem{Gliner}E. B. Gliner, \textit{Sov. Phys. JETP} \textbf{22}, 378 (1966).

\bibitem{Wald_book}R. Wald, \textit{General Relativity} (The University of Chicago Press, Chicago, 1984).

\bibitem{Dymnikova}I. G. Dymnikova, Gen. Relativ. Gravit. \textbf{24}, 235 (1992).

\bibitem{Dymnikova2} I. G. Dymnikova, Int. J. Mod. Phys. D \textbf{05}, 529 (1996).

\bibitem{Dymnikova3} I. G. Dymnikova, Int. J. Mod. Phys. D \textbf{12}, 1015 (2003). 

\bibitem{Bronnikov}K. A. Bronnikov, Phys. Rev. D \textbf{63}, 044005 (2001).

\bibitem{Hayward}S. A. Hayward, Phys. Rev. Lett. \textbf{96}, 031103 (2006).

\bibitem{Various_axial}A. Smailagic, E. Spallucci, Phys. Lett. B \textbf{688}, 82 (2010).

\bibitem{Various_axial2}L. Modesto, P. Nicolini, Phys. Rev. D \textbf{82}, 104035 (2010).

\bibitem{Various_axial3}C. Bambi, L. Modesto, Phys. Lett. B \textbf{721}, 329 (2013).

\bibitem{Various_axial4}B. Toshmatov, B. Ahmedov, A. Abdujabbarov, Z. Stuchlik, Phys. Rev. D \textbf{89}, 104017 (2014).

\bibitem{Various_axial5}M. Azreg-Ainou, Phys. Rev. D \textbf{90}, \ 064041 (2014).

\bibitem{Neves_Saa}J. C. S. Neves, A. Saa, Phys. Lett. B \textbf{734}, 44 (2014).

\bibitem{Neves}J. C. S. Neves, Int. J. Mod. Phys. A \textbf{32}, 1750112 (2017).

\bibitem{Bronnikov2} K. A. Bronnikov, V. N. Melnikov, H. Dehnen, Phys. Rev. D \textbf{68}, 024025 (2003).

\bibitem{Molina_Neves}C. Molina, J. C. S. Neves, Phys. Rev. D \textbf{82}, 044029 (2010).

\bibitem{Neves2}J. C. S. Neves, Phys. Rev. D \textbf{92}, 084015 (2015).

\bibitem{Hawking}S. W. Hawking, Commun. Math. Phys. \textbf{43}, 199 (1975).

\bibitem{Parikh}M. K. Parikh and F. Wilczek, Phys. Rev. Lett. \textbf{85}, 5042 (2000).

\bibitem{Agheben}M. Agheben, M. Nadalini, L. Vanzo, and S. Zerbini, JHEP \textbf{05}, 014 (2005).

\bibitem{Mann}R. Kerner and R. B. Mann, Phys. Rev. D \textbf{73}, 104010 (2006).

\bibitem{Mann2}R. Kerner and R. B. Mann, Class. Quant. Grav. \textbf{25}, 095014 (2008).

\bibitem{Ma}Meng-Sen Ma, and Ren Zhao, Class. Quantum Grav. \textbf{31}, 245014 (2014).

\bibitem{Bekenstein}J. D. Bekenstein, Phys. Rev. D \textbf{7}, 2333 (1973).

\bibitem{Tawfik}A. Tawfik, and A. Diab, Int. J. Mod. Phys. D \textbf{23}, 1430025 (2014).

\bibitem{Kaul}R. K. Kaul and P. Majumdar, Phys. Rev. Lett. \textbf{84}, 5255 (2000).

\bibitem{Various_GUP}Y.-W. Kim and Y.-J. Park, Phys. Lett. B \textbf{655}, 172 (2007).

\bibitem{Various_GUP2}M. Yoon, J. Ha, and W. Kim, Phys. Rev. D \textbf{76}, 047501 (2007).

\bibitem{Various_GUP3}K. Nouicer, Phys.Lett. B \textbf{646}, 63 (2007).

\bibitem{Anacleto}M. A. Anacleto, F. A. Brito, E. Passos, Phys. Lett. B \textbf{749}, 181 (2015).

\bibitem{Casadio}R. Casadio, P. Nicolini, R. da Rocha, \textit{GUP Hawking fermions from MGD black holes} [arXiv:1709.09704].

\bibitem{Anacleto2}M. A. Anacleto, D. Bazeia, F. A. Brito, J. C. Mota-Silva, Adv. High Energy Phys. \textbf{2016}, 8465759 (2016).

\bibitem{Beato}E. Ayon-Beato, A. Garcia, Phys. Lett. B \textbf{493}, 149 (2000) .

\bibitem{Thermodynamics}Y. S. Myung, Y. W. Kim, Y. J. Park, Gen. Rel. Grav. \textbf{41}, 1051 (2009).

\bibitem{Thermodynamics2}J. Man, H. Cheng, Gen. Rel. Grav. \textbf{46}, 1660 (2014).

\bibitem{Thermodynamics3}N. Breton, Gen. Rel. Grav. \textbf{37}, 643 (2005).

\bibitem{Thermodynamics4}M. Akbar, N. Salem, S. A. Hussein, Chin. Phys. Lett  \textbf{29}, 070401 (2012).

\bibitem{Thermodynamics5}L. Balart, Phys. Lett. B \textbf{687}, 280 (2010).

\bibitem{Thermodynamics6}Y. S. Myung, Y. W. Kim, Y. J. Park, Phys. Lett. B \textbf{659}, 832 (2008).

\bibitem{Thermodynamics7}W. Kim, H. Shin, M. Yoon, J. Korean Phys. Soc. \textbf{53}, 1791 (2008).

\bibitem{Tharanath}R. Tharanath, J. Suresh, V. C. Kuriakose, Gen. Rel. Grav. \textbf{47}, 46 (2015).

\bibitem{Sharif}M. Sharif, W. Javed, Can. J. Phys. \textbf{89}, 1027 (2011).

\bibitem{Olmo}G. J. Olmo, D. Rubiera-Garcia, A. Sanchez-Puente, Phys. Rev. D \textbf{92}, 044047 (2015).

\bibitem{Wald}R. Wald, Living Rev. Rel. \textbf{4}, \ 6 (2001).

\bibitem{Nicolini}P. Nicolini, A. Smailagic and E. Spallucci, Phys. Lett. B \textbf{632}, 547 (2006).

\bibitem{applications}Z. Li, C. Bambi, JCAP\textbf{1401}, 041 (2014).

\bibitem{applications2}C.Bambi, Phys. Lett. B \textbf{730}, 59 (2014).

\bibitem{applications3}S. G. Ghosh, P. Sheoran, M. Amir, Phys. Rev. D \textbf{90}, 103006 (2014).

\bibitem{applications4}M. Amir, S. G. Ghosh, JHEP \textbf{1507}, 015 \ (2015).

\bibitem{applications5}A. Flachi, J. P. S. Lemos, Phys. Rev. D \textbf{87}, 024034 (2013).

\bibitem{applications6}J. Li, K. Lin, N. Yang, Eur. Phys. J. C \textbf{75}, 131 (2015).

\bibitem{applications7}K. A. Bronnikov, R. A. Konoplya, A. Zhidenko, Phys. Rev. D \textbf{86}, 024028 (2012).

\bibitem{applications8}B. Toshmatov, A. Abdujabbarov, Z. Stuchlík, B. Ahmedov, Phys. Rev. D \textbf{91}, 083008 (2015).

\bibitem{applications9}E. F. Eiroa, C. M. Sendra, Class. Quant. Grav. \textbf{28}, 085008 (2011).

\bibitem{applications10}V. Santos, R. V. Maluf, and C. A. S. Almeida, Phys. Rev. D \textbf{93}, 084047 (2016).

\bibitem{Vagenas}A. J. M. Medved, and E. C. Vagenas, Phys. Rev. D \textbf{70}, 124021 (2004).

\bibitem{Das}S. Das, E. C. Vagenas, Phys. Rev. Lett. \textbf{101}, 221301 (2008).

\bibitem{Vagenas2}F. Scardigli, R. Casadio, Eur. Phys. J. C \textbf{75}, 425 (2015).

\end{thebibliography}
\end{document}